\documentstyle[eqsecnum,pre,preprint,aps,epsfig]{revtex}

\tightenlines

\newcommand{\bq}{\begin{equation}}
\newcommand{\eq}{\end{equation}}
\newcommand{\bqa}{\begin{eqnarray}}
\newcommand{\eqa}{\end{eqnarray}}

\begin{document}

\draft
\preprint{PM/01-18}

\title{Top Quark Production at TeV Energies as a Potential \\
 SUSY Detector
\footnote{Partially
supported by EU contract  HPRN-CT-2000-00149.}}

\author{M. Beccaria$^{a,b}$, S. Prelovsek$^{c}$, 
F.M. Renard$^d$ and C. Verzegnassi$^{e, f}$ \\
\vspace{0.4cm} 
}

\address{
$^a$Dipartimento di Fisica, Universit\`a di 
Lecce \\
Via Arnesano, 73100 Lecce, Italy.\\
\vspace{0.2cm}  
$^b$INFN, Sezione di Lecce\\
\vspace{0.2cm} 
$^c$ Physics Department, University of Ljubljana and \\
Institute J. Stefan, Ljubljana, Slovenia\\
\vspace{0.2cm}
$^d$ Physique
Math\'{e}matique et Th\'{e}orique, UMR 5825\\
Universit\'{e} Montpellier
II,  F-34095 Montpellier Cedex 5.\hspace{2.2cm}\\
\vspace{0.2cm} 
$^e$
Dipartimento di Fisica Teorica, Universit\`a di Trieste, \\
Strada Costiera
 14, Miramare (Trieste) \\
\vspace{0.2cm} 
$^f$ INFN, Sezione di Trieste\\
}

\maketitle

\begin{abstract}

We consider the process of top-antitop production from
electron-positron annihilation, for c. m. energies in the few TeV
regime, in the MSSM theoretical framework. We show that, at the one
loop level, the \underline{slopes} of a number of observable
quantities in an energy region around 3 TeV are only dependent on
$\tan\beta$. Under optimal experimental conditions, a combined
measurement of slopes might identify $\tan\beta$ values in a range
$\tan\beta< 2$, $\tan\beta>20$ with acceptable precision.

\end{abstract}


\section{Introduction.} 

The existence of large electroweak Sudakov logarithms \cite{Sudakov}
in four-fermion
processes at the one-loop level, for c.m. energies in the TeV range
\cite{KDD,log},
and the subsequent efforts for providing a full resummation of the
relevant terms \cite{Melles,sum}, 
have been the subject of recent theoretical motivated
discussions and effort, whose final elaboration is still in progress
within the theoretical SM framework.  More recently, the extension of
this kind of analysis at one loop within the MSSM, 
for final SM
massless fermion \cite{slog}
and top quark pair production \cite{susytoplog}, has been also
accomplished. Finally, very recently, an analysis of the role of the
so calles ``$\theta$-independent'' and ``$\theta$-dependent'' Sudakov
effects at one loop has been presented \cite{unsud}.  
These two terms are, briefly,
those for which a clean resummation prescription exists, and those for
which this conclusion does not seem to be valid \cite{sum}. 
In the t'Hooft
$\xi=1$ gauge, as shown in \cite{unsud}, 
the latter terms are subleading
linear logarithms in the c.m. energy $\sqrt{q^2}$, originated by a
$\theta$-dependent ($\theta$ is the c.m. scattering angle) component
of the $W$ (and, to a much smaller extent, $Z$) boxes.

One of the conclusions of \cite{susytoplog} is that the process of
\underline{top} pairs production from electron-positron annihilation
at c.m. energies in the few TeV range, i.e. those that will be
explored by the future CERN CLIC collider \cite{CLIC}, 
is particularly lucky for
what concerns the validity of a SM calculation. Owing to the weak
isospin characterization of the final state, in fact, the non
resummable $\theta$-dependent contribution turns out to be quite
small \cite{unsud}, 
so that only the resummable $\theta$-independent component
survives in the SM, making a clean theoretical estimate of the various
observables already (in principle) available. Note that similar
conclusions would not apply, for instance, for bottom production, due
to its opposite isospin value.

The fact that virtual SM effects are fully under control raises the
interest of considering the role of possible virtual SUSY effects of
Sudakov origin in top production at CLIC energies. On very general
grounds, we shall say that to be considered as potentially
``interesting'', such effects should be of an ``intelligent'' type. By
this we mean that they should be visible, and therefore ``sufficiently
large'' with respect to the experimental accuracy at the computable
one-loop level. On the other hand they should also be ``sufficiently
small'' at the same level, not to spoil its presumed perturbative
validity (given the fact that a two loop calculation of SUSY virtual
effects seems to us, at least at the moment, unrealistic for this
process.)

It was shown in \cite{susytoplog} that the leading Sudakov SUSY effect at one
loop is only of linear logarithmic kind (in the SM, also quadratic
logarithms appear), and $\theta$-independent. It is only produced by
final vertices, an contains a component of ``massless quark'' kind and
one of ``massive top'' Yukawa origin that strongly depends on
$\tan\beta$. Its numerical effect can vary from a few percent to more
than ten percent, strongly depending on $\tan\beta$, being apparently
still sensitive to relatively large $\tan\beta$ values $>10$. In
~\cite{susytoplog}, it was suggested that this fact might be used to try to
fix the value of $\tan\beta$ from a combined analysis of the value of
several observables at \underline{fixed} energy, e.g. around the
proposed CLIC ``optimal'' value $\sqrt{q^2} = 3~ TeV$. The conclusion
of that Reference was that a deeper investigation of this possibility
would have followed.  

The aim of this short paper is precisely that of
performing the aforementioned investigation. Our proposal will be that
of considering, rather than measurements at fixed energies, variations
of observables (slopes) with energy around a chosen specially
interesting (e.g. 3 TeV) energy value. We shall show that the only
unknown quantities in the coefficients of the various slopes are
functions of $\tan\beta$ alone. All the other MSSM parameters can be
incorporated asymtotically into terms that either vanish or remain
constant, thus disappearing in the slope. We shall then consider a
reasonable experimental setup and try to conclude that it might be
possible to identify $\tan\beta$ with ``decent'' precision (e.g. to
better than a relative fifty percent) up to relatively large $\tan\beta$
values. This should be compared and combined with other interesting
recently proposed $\tan\beta$ detection techniques~\cite{djouadi}.  

Technically
speaking, this short paper is organized as follows. In Section II we
shall present our one-loop derivation of the SUSY slopes in the few
TeV region. In Section III, we shall propose our ``data simulation''
and $\tan\beta$ identification, and present a few general concluding
remarks in the final Section IV.

\section{SUSY Sudakov logarithms in the TeV region}

A first treatment of top production at one loop in the MSSM at TeV
energies has been already given \cite{susytoplog}, 
and all the relevant formulae for
observables can be found there in Section IV. For the specific
purposes of this paper, we shall rewrite them here in a form where the
SM component of all observables has been considered as a given
perfectly known theoretical input. This statement requires the
following precise explanation. At few TeV energies, it has been argued
very recently \cite{unsud}
that the one-loop theoretical description in the SM is
often in trouble, due to the presence of large and opposite
``$\theta$-independent'' and ``$\theta$-dependent'' Sudakov terms. The
possible way out would be represented by a resummation of the
separately large logarithms. Unfortunately, a clean resummation
prescription at the moment only seems to exist for the so called
``$\theta$-independent'' terms, and not for the ``$\theta$-dependent''
ones \cite{sum}, 
so that a completely satisfactory set of theoretical predictions
in the few TeV (CLIC) energy range must still be provided.

A remarkable exception to this negative statement is represented by
top pair production. Here, for reasons that are simply connected with
the top weak isospin assignment \cite{unsud}, 
the $\theta$-dependent terms are
strongly depressed at one loop, so that their resummation does not
seem necessary.  A partial resummation of the separate
$\theta$-independent terms would thus guarantee a fully satisfactory
theoretical prediction for the SM component of the process, leaving
the SUSY component as the only quantity to be investigated.

Following this attitude, we shall thus rewrite all the relevant
formulae of \cite{susytoplog} indicating with the ``SM'' label the
(supposedly perfectly known) SM component. For our purposes, we shall
need the asymptotic Sudakov expansion of the various quantities. 
The latter
contains as the leading term a $\theta$-independent, linear
logarithm. This is, as we said in the Introduction, the sum of a
``massless'' and a ``massive'' component, whose origin is due to the
vertex diagrams shown in Fig.(1). In the limit when the c.m. energy
$\sqrt{q^2}$ becomes very large, they produce the overall leading
logarithmic SUSY Sudakov terms listed in the following equations for
photon or $Z$ exchanges:

\bqa
\Gamma^{\gamma}_{\mu}(\mbox{MSSM},~ \mbox{massive})&\to&
{e\alpha\over12\pi M^2_Ws^2_W}\ \ln\frac{ q^2}{M^2}
\{m^2_t(1+\cot^2\beta)[(\gamma_{\mu}P_L)+2(\gamma_{\mu}P_R)]\nonumber\\
&&
+m^2_b(1+\tan^2\beta)(\gamma_{\mu}P_L)\}
\label{MSSMgm}\eqa

\bqa
\Gamma^{Z}_{\mu}(\mbox{MSSM},~ \mbox{massive})&\to&
{e\alpha\over48\pi M^2_Ws^3_Wc_W}\ \ln\frac {q^2}{M^2}
\{(3-4s^2_W)m^2_t(1+\cot^2\beta)(\gamma_{\mu}P_L)\nonumber\\
&&
-8s^2_Wm^2_t(1+\cot^2\beta)(\gamma_{\mu}P_R)
+(3-4s^2_W)m^2_b(1+\tan^2\beta)(\gamma_{\mu}P_L)\}
\label{MSSMZm}\eqa

\bq
\Gamma^{\gamma}_{\mu}(\mbox{MSSM},~ \mbox{massless}) \to
{e\alpha\over12\pi  s^2_W c^2_W}\ \ln \frac{q^2}{M^2}
\left\{ \left(3-\frac{26}{9}s_W^2\right) P_L + \frac{16}{9}s_W^2 P_R \right\}
\label{MSSMgms}\eq

\bq
\Gamma^{Z}_{\mu}(\mbox{MSSM},~ \mbox{massless}) \to
{e\alpha\over32\pi s^3_W c^3_W}\ \ln\frac{ q^2}{M^2}
\left\{ \left(3-\frac{62}{9}s_W^2+\frac{104}{27}s_W^4\right) P_L - 
\frac{64}{27}s_W^4 P_R \right\}
\label{MSSMZms}\eq
where $P_{L, R} = \frac 1 2 (1\mp\gamma_5)$.

One sees that, in the leading term, the only SUSY parameters that
appear are $\tan\beta$ and an overall common unknown ``SUSY scale''
$M$ which we only assumed to be ``reasonably'' smaller than the energy
value $\sqrt{q^2}=3$ TeV in which we are interested in this work. Of
course, this assumption might be wrong and heavier SUSY masses might
turn out to be produced. In that case, our ``asymptotic'' expansions
would still be valid, obviously in a suitably larger energy range.

Starting from Eqs.(\ref{MSSMgm}),(\ref{MSSMZm}),
it is a straightforward task to derive the
leading SUSY Sudakov contributions to the various observables. We
write here, in the previously discussed spirit, the expressions that
will be relevant for our purposes, considering, for simplicity, first
a set of observables where the final top quark helicity is not
measured and secondly a set of 4 observables where it is. The chosen
quantities are $\sigma_t$ (the cross section for top pair production),
$A_{FB, t}$ and $A_{LR, t}$ (the forward-backward and longitudinal
polarization asymmetries) and $A_t$ (the polarized forward-backward
asymmetry) in the first set.  In the second one, we have considered
$H_t$ (the averaged top helicity), $H_{FB, t}$ (its forward-backward
asymmetry), $H^{LR}_t$ (the averaged polarized top helicity) and
$H_{t, FB}^{LR}$ (its forward-backward asymmetry), and all these
quantities are defined in detail in the Appendix B of \cite{susytoplog}.  For
the chosen observables we obtain the following asymptotic expansions:

\bqa
\sigma_{t}&=&\sigma^{SM}_{t}\{1+{\alpha\over4\pi}\{{ (4.44\,N+11.09
)\,\ln{q^2\over\mu^2}-10.09\, \ln{q^2\over M^2}\}
+F_{\sigma_t}(\tan\beta) \ln{q^2\over M'^2}}
\label{sigtt}\eqa

$$
F_{\sigma_t}(t) = \frac\alpha{4\pi} (-29\ t^{-2}-0.0084\ t^2 -14),
$$

\bqa
A_{FB,t}&=&A^{SM}_{FB,t}+{\alpha\over4\pi}\{{ (0.22\,N+1.29
)\,\ln{q^2\over\mu^2}-0.23\, \ln{q^2\over M^2}\}
+F_{A_{FB,t}}(\tan\beta) \ln{q^2\over M'^2}}
 \ .
\label{AFBt}\eqa

$$
F_{A_{FB,t}}(t) = \frac{\alpha}{4\pi}(1.2\ t^{-2}-0.00082\ t^2 +0.62),
$$

\bqa
A_{LR,t}&=&A^{SM}_{LR,t}+{\alpha\over4\pi}\{{ (1.03\,N+5.95
)\,\ln{q^2\over\mu^2}-4.03\, \ln{q^2\over M^2}\}
+F_{A_{LR,t}}(\tan\beta) \ln{q^2\over M'^2}}
\label{ALRt}\eqa

$$
F_{A_{LR,t}}(t) = \frac{\alpha}{4\pi}(7.7\ t^{-2}-0.0051\ t^2 +3.8),
$$

\bqa
A_{t}&=&A^{SM}_{t}+{\alpha\over4\pi}\{{ (0.91\,N+5.25
)\,\ln{q^2\over\mu^2}-3.20\, \ln{q^2\over M^2}\}
+F_{A_{t}}(\tan\beta) \ln{q^2\over M'^2}},
\label{AFBpt}\eqa

$$
F_{A_t}(t) = \frac{\alpha}{4\pi}(7.5\ t^{-2}-0.0049\ t^2 +3.7),
$$

\bqa
H_t&=&H^{SM}_t+{\alpha\over4\pi}\{{ (-1.21\,N-7.00
)\,\ln{q^2\over\mu^2}+4.27\, \ln{q^2\over M^2}\}
+F_{H_t}(\tan\beta) \ln{q^2\over M'^2}},
\label{Ht}\eqa

$$
F_{H_t}(t) = \frac{\alpha}{4\pi}( -9.9\ t^{-2}+0.0066\ t^2 -5),
$$

\bqa
H_{t,FB}&=&H^{SM}_{t,FB}+{\alpha\over4\pi}\{{ (-0.77\,N-4.46
)\,\ln{q^2\over\mu^2}+3.02\, \ln{q^2\over M^2}\}
+F_{H_{t,FB}}(\tan\beta) \ln{q^2\over M'^2}},
\label{Hfbt}\eqa

$$
F_{H_{t,FB}}(t) = \frac{\alpha}{4\pi}(-5.7\ t^{-2}+0.0038\ t^2 -2.9),
$$

\bqa
H^{LR}_t&=&H^{LR,SM}_t+{\alpha\over4\pi}\{{ (-0.30\,N-1.71
)\,\ln{q^2\over\mu^2}+0.31\, \ln{q^2\over M^2}\}
+F_{H^{LR}_t}(\tan\beta) \ln{q^2\over M'^2}},
\label{HLRt}\eqa

$$
F_{H^{LR}_t}(t) = \frac{\alpha}{4\pi}(-1.6\ t^{-2}+0.0011\ t^2 -0.82),
$$

\bqa
H^{LR}_{t,FB}&=&H^{LR,SM}_{t,FB}+{\alpha\over4\pi}\{{
 (0.)\,\ln{q^2\over\mu^2}+(0.)\, \ln{q^2\over M^2}\}
+F_{H^{LR}_{t,FB}}(\tan\beta)\ln{q^2\over M'^2}},
\label{HLRFBt}\eqa

$$
F_{H^{LR}_{t,FB}}(t) = \mbox{constant}
$$

In the previous equations, we have also listed in the first term,
for sake of completeness, the SUSY asymptotic linear logarithm of RG
origin. This contributes a universal term of self-energy origin, where
no SUSY parameters (except a SUSY scale) appear. In our procedure we
shall add this RG logarithm to that of SM origin, and consider it as
an uninteresting part of the ``Non SUSY Sudakov'' structure, 
that will be treated as a \underline{known} contribution in 
our ``slopes-based'' procedure. 
The second term is the massless
Sudakov term (i.e. the one that would have been obtained for $u$, $c$
pair production) and the third one contains the massive Sudakov term
as a function of $\tan\beta$.

One notices, as stressed in \cite{susytoplog}, a strong $\tan\beta$
dependence in some observable (particularly $\sigma_t$) that might
possibly be exploited to perform a determination of this
parameter. With this purpose, we have examined the above possibility with
some caution, in a way that shall now illustrate.

Clearly, if the logarithmic term were the only relevant one in the
SUSY component of the observables, a determination of $\tan\beta$
might proceed in principle via a fit of the various observables at a
fixed chosen energy.  Quite generally, in an asymptotic expansion like
the one that we are assuming, there will be extra non leading
contributions, in particular constant terms and terms that vanish (at
least as $1/q^2$) asymptotically.  We assumed consistently within our
philosophy that the latter ones can be safely neglected and
concentrated our attention on possible constant quantities. Our
approach was that of computing \underline{exactly} the contributions
to the various observables from the considered SUSY vertices, and to
try to fit the numerical results with an expression of the kind
$(F_1\cot^2\beta + F_2\tan^2\beta+F_3)\log q^2 + G$, as discussed
in the forthcoming Section.

\section{Numerical Validity of the Asymptotic Expansion and 
a Possible Procedure for the Determination of $\tan\beta$}

To check the validity of the Sudakov asymptotic expansion we
have computed the relevant complete one loop SUSY effects in the MSSM 
by evaluating without approximations all the diagrams that are 
not vanishing in the large energy limit, considering a (ideal) situation
in which at least \underline{some} of the SUSY parameters have been
measured with reasonable precision while other ones remain possibly
still undetermined.

The large set of free parameters of the MSSM has been 
chosen for a first approach 
as follows:  we have assumed the Grand Unification relation 
$M_1 = \frac 5 3 \sin^2\theta_W M_2$ 
between the $U(1)$ and $SU(2)$ gaugino parameters;
we have fixed the mass of the lightest chargino at 200 GeV,  
the $\mu$ parameter at 
500 GeV and the mass of the 
CP odd neutral Higgs boson at 300 GeV; 
we have considered a mixed sfermion sector
characterized by a sfermion mass scale $M_S$ = 300 GeV.

For each observable we have 
attempted a simple parametrization of the SUSY effect 
of the form 
\bq
F \log\frac{q^2}{M_Z^2} + G
\eq
where $F$ and $G$ depend on the observable and in principle also on all the 
model parameters. In fact, we know that if we are in an energy range where 
the Sudakov expansion is holding than the coefficient $F$ depends on 
$\tan\beta$ only and admits the simple parametrization 
\bq
\label{functional}
F \equiv F(\tan\beta) = F_1\cot^2\beta+ F_2\tan^2\beta+F_3
\eq
with $F_1$, $F_2$ and $F_3$ in agreement with Eqs.~(\ref{sigtt}-\ref{HLRt}).

In Fig.(\ref{sigmat}) we plot the percentual relative SUSY effect 
in $\sigma_t$ computed with the above mentioned set of MSSM parameters and 
by choosing $\tan\beta=2.0$. On the logarithmic scale it is easy to 
recognize the asymptotic linear logarithmic behaviour that sets in 
at energies beyond the threshold $\sqrt{q^2}\simeq 3\ {\rm TeV}$.
A similar behaviour can be seen in all the other observables.

We then have repeated the analysis for several values of $\tan\beta$ and 
have tried to determine the dependence of $F$ and $G$ on that parameter.
The results are shown in Fig.(\ref{fit}). Circles and diamonds are the data 
points that have been 
obtained by a fit performed under the conservative cut 
$\sqrt{q^2} > 3\ \rm TeV$. 
The solid and dashed lines have been obtained by fitting the data with 
the functional form Eq.~(\ref{functional}). The match is quite good and the 
crucial point is that \underline{$F_1$, $F_2$ and $F_3$
 agree with the analytical 
Sudakov expansion} to a few percent. As such, they are actually
known numerical constants independent on other model parameters like 
gaugino mass parameters or other SUSY scales.
In fact, our choice for the set of MSSM parameters is
``reasonable'' lacking
more detailed experimental information and it permits to check 
if the typical energy at which the Sudakov expansion starts to be reliable
is in the CLIC reach.
If some MSSM parameter is varied, for instance assuming a heavier lightest
chargino, then our conclusions are still valid asymptotically. Although 
we did not 
perform an exhaustive investigation of the full parameter case, 
in several sensible cases $\sqrt{q^2} > 3\ \rm
 TeV$ appears to be a safe threshold for the expansion.

Is it possible to exploit this simplification in the high energy limit 
to determine the value of $\tan\beta$ ? To address this question, 
we define the relative effect on the observable 
${\cal O}_n$ as the ratio
\bq
\epsilon_n(q^2) = \frac{{\cal O}_n(q^2)-{\cal O}^{SM}_n(q^2)}{{\cal O}^{SM}_n(q^2)} = 
F_n\log\frac{q^2}{M^2_Z} + G_n
\eq
and denote by $\sigma_n(q^2)$ the experimental error on
 $\epsilon_n(q^2)$, assuming ${\cal O}^{SM}_n$ perfectly known.

We then suppose that a set of $N$ independent measurements is available 
at c.m. energies $\sqrt{q_1^2}, \sqrt{q_2^2}, \dots,\sqrt{q_N^2}$. 
Differences with respect to the measurement at lowest energy
\bq
\delta_{n, i} = \epsilon_n(q_i^2)-\epsilon_n(q_1^2)
\eq
do not contain the constant $G_n$ (and the SUSY mass scales $M$, $M'$
hidden in it)
and provide direct access to 
$\tan\beta$ through $F$. In fact, 
for each set of explicit measurements $\{\delta_n(q_i^2)\}$, 
the optimal value of $\tan\beta$ 
is determined by minimizing the $\chi^2$ sum
\bq
\chi^2 = \sum_{i=1}^N\sum_{n=1}^{N_{\cal O}}
\frac{(F_n \log\frac{q^2_{i+1}}{q^2_1} - \delta_{n, i})^2}{4\sigma_{n, i}^2}
\eq
where $\delta_{n, i} \equiv \delta_n(q^2_i)$ and $\sigma_{n, i} \equiv \sigma_n(q_i^2)$.
We make the usual assumption that $\delta_{n, i}$ is a normal Gaussian 
random variable distributed around the value
\bq
F_n(\tan\beta^*) \log\frac{q^2_{i+1}}{q^2_1}
\eq
with standard deviation $2\sigma_{n,i}$. Hence, $\beta^*$ is the 
unknown true value. After linearization around $\tan\beta=\tan\beta^*$, 
minimization of $\chi^2$ provides the best estimate of $\tan\beta$ 
that is  also a Gaussian random variable.
Its mean is of course $\tan\beta^*$ and 
the standard deviation $\delta\tan\beta$ 
is given by the condition $\Delta\chi^2=1$ 
\bq
\delta\tan\beta = 2\left(\sum_{n, i} \left(\frac{F_n'(\tan\beta^*) 
\log\frac{q_{i+1}^2}{q_1^2}}
{\sigma_{n,i}}\right)^2\right)^{-1/2}
\eq
If we simplify the discussion by assuming $\sigma_{n, i} \equiv \sigma$, this
formula reduces to 
\bq
\delta\tan\beta = 2\sigma\left(\sum_n F_n'(\tan\beta^*)^2\right)^{-1/2}
\left(\sum_i \log^2\frac{q_{i+1}^2}{q_1^2} \right)^{-1/2}
\eq
The function 
\bq
\tau(\tan\beta) = \left(\sum_n F_n'(\tan\beta)^2\right)^{-1/2}
\eq
measures the dependence of the slope of SUSY effects on $\tan\beta$. It is 
shown in Fig.(\ref{tau}) for three possible choices: (i)
only $\sigma_t$, (ii) the four non - helicity observables
$\sigma_t$, $A_{FB, t}$, $A_{LR, t}$ and $A_t$, (iii) the non-helicity 
observables and the three helicity observables 
$H_t$, $H_{FB, t}$ and $H^{LR}_t$.

In the best case (iii), it is strongly peaked around $\tan\beta = 8$
and the combination of the various observables, 
especially $\sigma_t$, $A_{LR, t}$, $A_t$ and $H_t$
(the ones with larger $\cot^2\beta$ coefficient) is crucial to keep 
the function $\tau(\tan\beta)$ as small as possible.

To understand the consequences of the shape of $\tau$, we plot in 
Fig.(\ref{error}) the relative error $\delta\tan\beta/\tan\beta$ 
computed under the optimistic assumption of a relative accuracy 
(absolute for $\sigma_t$) equal to 1\% for all the seven observables.
The three curves correspond to the assumption that 
independent 
measurements for each observable are available at $N=5$, 10 or 20  
equally spaced c.m. energies  between 2 TeV and 6 TeV.
Of course, different curves associated to pairs $(N, \sigma)$ actually
depend only on the combination $\sigma/\sqrt{N}$.
We also show horizontal dashed lines corresponding to relative errors
equal to 1 and 0.5. As one can see in the Figure, values in the range
\bq
\tan\beta < 2, \qquad \tan\beta > 20
\eq
can be detected with $N=10$ c.m. energy values
with a relative error smaller than 50\%, that we consider
qualitatively as a ``decent'' accuracy. Obviously, if a higher experimental
precision (e.g. a few permille in $\sigma_t$) were achievable, the same
result could be obtained with a smaller number ($N\simeq 3$) of independent
energy measurements.

\section{Conclusions}

We have discussed in this paper the possibility of determining the crucial 
MSSM parameter $\tan\beta$ via measurements of the slopes of a number
of experimental observables in the process of top-antitop production from 
electron-positron annihilation in the CLIC regime. 
We have assumed an ``ideal'' situation in which some information on the 
SUSY parameters already exists, and we have fixed some of them to values
that appear to us reasonable. Of course, this values might be different than
those which (hopefully) will be determined in future measurements. The point
remains, though, that independently on this ``details'' the \underline{slopes}
of the observables will be \underline{only} dependent on $\tan\beta$.
Our results show that,
in principle, under expected reasonable experimental conditions, it would
be possible to derive ``decent'' informations on $\tan\beta$ in two ranges,
i.e. $\tan\beta < 2$ and $\tan\beta > 20$. This seems to us an interesting
possibility, particularly for what concerns the second large $\tan\beta$
range. In this case, to our knowledge, the realistic possibilities of
measuring $\tan\beta$ are rather
restricted and not simple, as exhaustively discussed in a very recent 
paper~\cite{djouadi}. 
In fact, a measurement of $\tan\beta$ is practically impossible from chargino
or neutralino production when $\tan\beta>10$ since the effects depend on 
$\cos2\beta$ that becomes flat for $\beta\to \pi/2$. It could be achieved 
in the associated productions $e^+e^-\to h\tilde\tau\tilde\tau$ or 
$e^+e^-\to A\bar{b}b$ (with $h$ and $A$ being the CP even and odd Higgs 
bosons), but only for very large $\tan\beta$ values ($\sim 50 $).
Our proposal might represent an alternative independent determination,
to be possibly combined with other methods, either already proposed
or to be suggested in future studies.

\begin{figure}[htb]
\vspace*{3cm}
\[
\epsfig{file=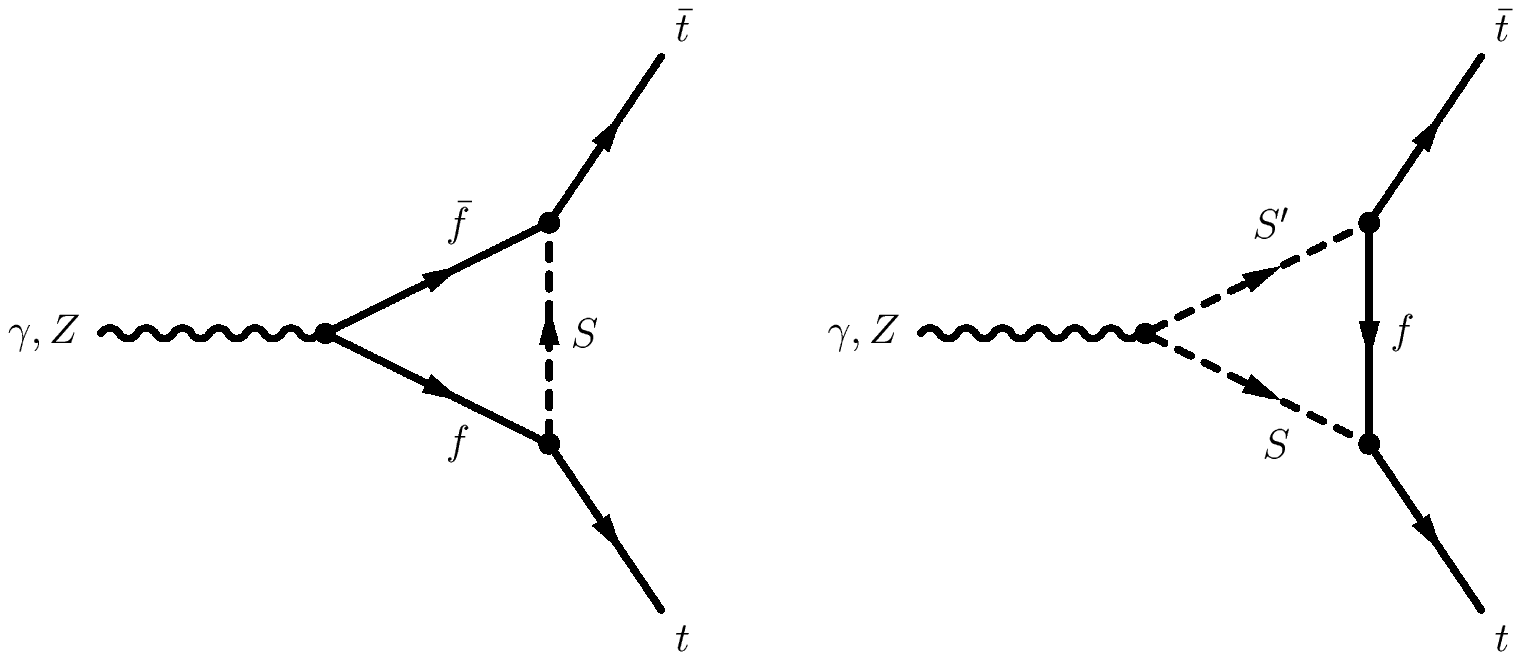,height=5cm}
\]
\vspace*{1.5cm}
\label{figS}
\end{figure}\begin{figure}[p]
\vspace*{-2cm}
\[
\epsfig{file=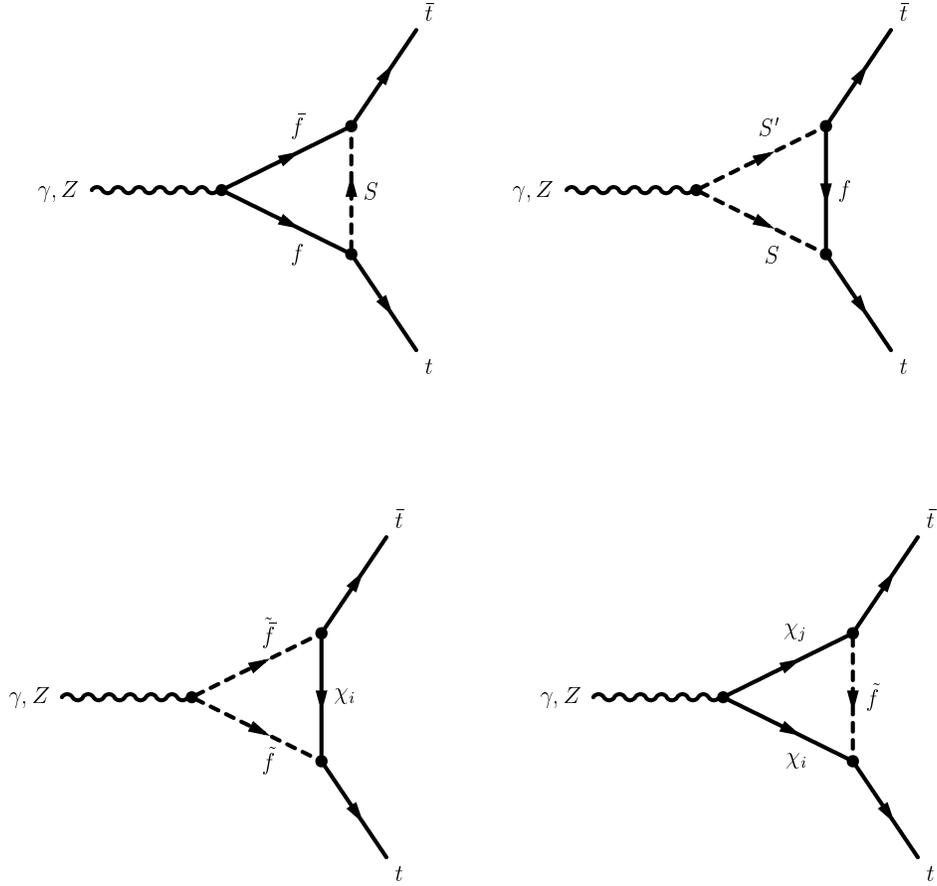,height=5cm}
\]
\vspace*{1.5cm}
\caption[2]{Triangle diagrams with SUSY Higgs and with SUSY
partners contributing 
to the asymptotic logarithmic behaviour in the energy; 
$f$ represent $t$ or $b$ quarks; $S$ 
represent charged or neutral Higgs bosons $H^{\pm}$,
$A^0$, $H^0$, $h^0$ or Goldstone $G^0$; $\tilde f$ represent
stop or sbottom states; $\chi$ represent charginos or neutralinos.
The arrow corresponds to the momentum flow of the indicated particle.}
\label{figsusy}
\end{figure}

\begin{figure}[htb]
\vspace*{2cm}
\[
\epsfig{file=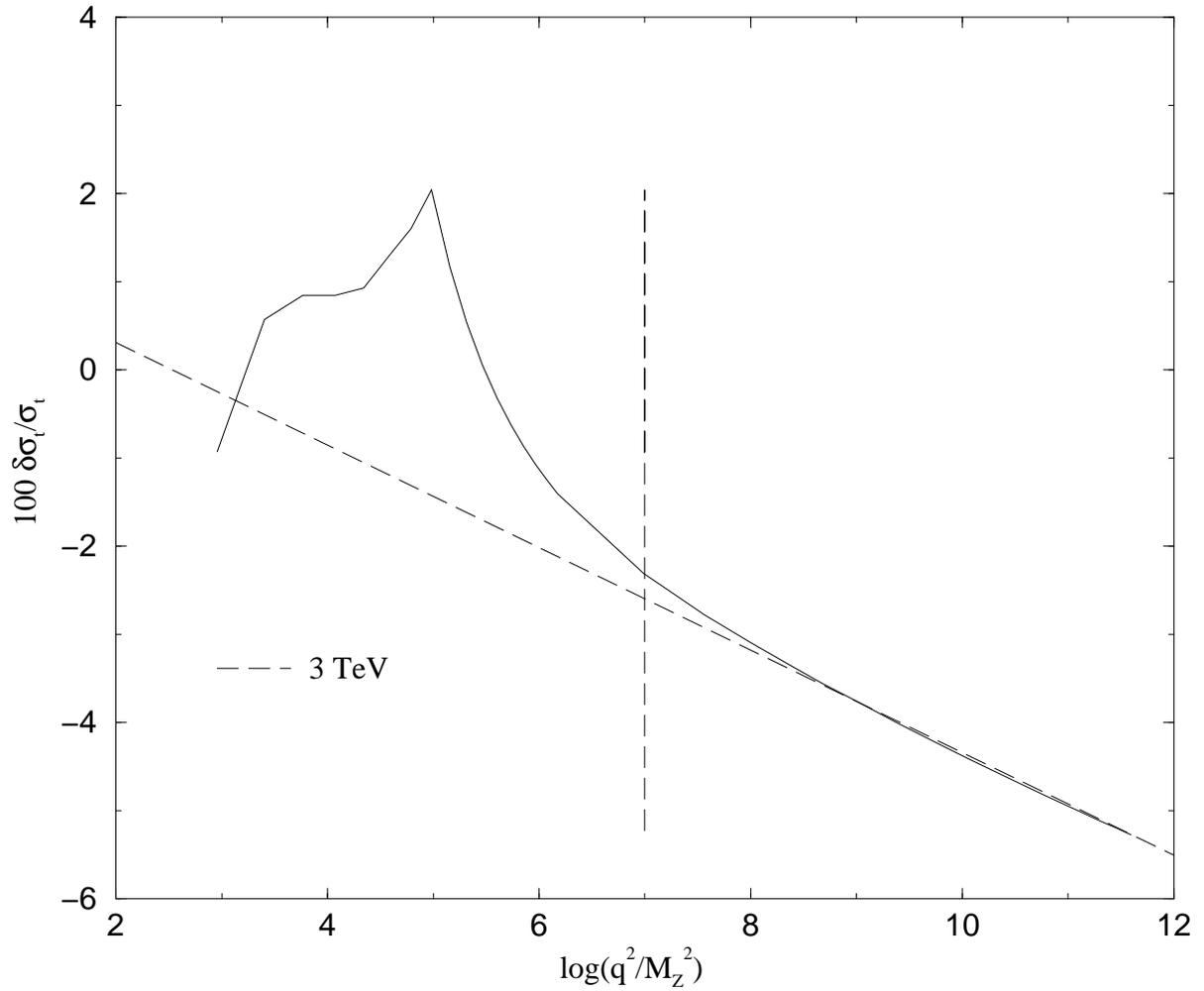,height=16cm,angle=-90}
\]
\vspace*{1cm}
\caption[1]{
Plot of $\sigma_t$. The set of MSSM parameters is fully described in 
Section III.
The vertical dashed line marks the CLIC 3 TeV energy. The oblique dashed 
line is 
a linear logarithmic fit of the curve in the high energy region.
}
\label{sigmat}
\end{figure}
\newpage

\begin{figure}[htb]
\vspace*{2cm}
\[
\epsfig{file=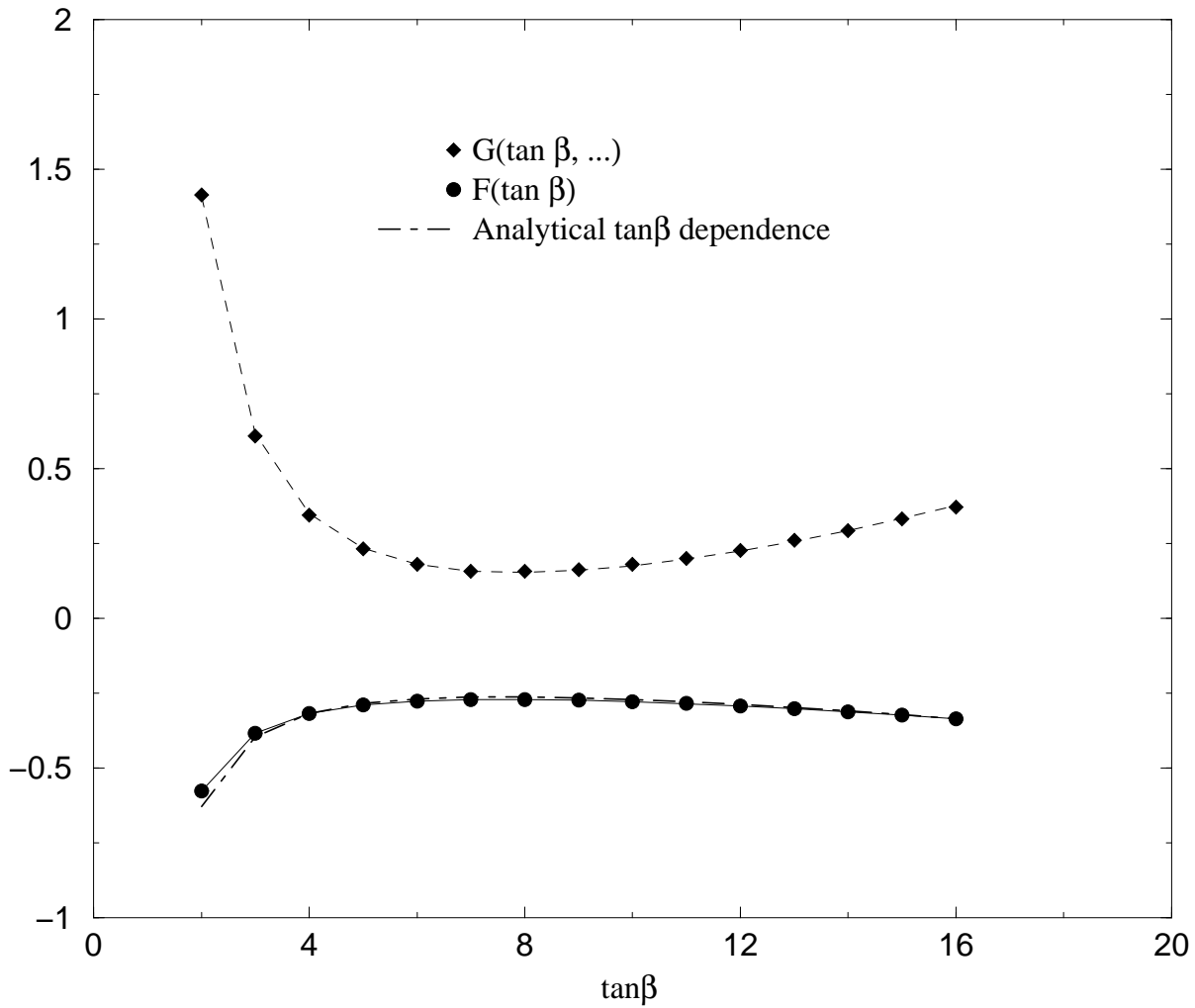,height=16cm,angle=-90}
\]
\vspace*{1cm}
\caption[1]{
Logarithmic fit of the SUSY effects in $\sigma_t$ and $\tan\beta$ dependence.
The data points (circles and diamonds) are the result from a logaritmic fit
of the relative SUSY effects in $\sigma_t$ as in Fig.~(\ref{sigmat}).
The solid and dashed lines are nonlinear fits of the functional form 
$F_1\cot^2\beta+F_2\tan^2\beta+F_3$.
As we explained, the results for $F$ confirm the validity of
the analytical Sudakov expansion that is also shown in the Figure (dot-dashed
line).
It is interesting to remark that the same functional dependence on $\tan\beta$
seems to work well also for the constant $G$ (for a given choice of the other
MSSM parameters).
}
\label{fit}
\end{figure}
\newpage

\begin{figure}[htb]
\vspace*{2cm}
\[
\epsfig{file=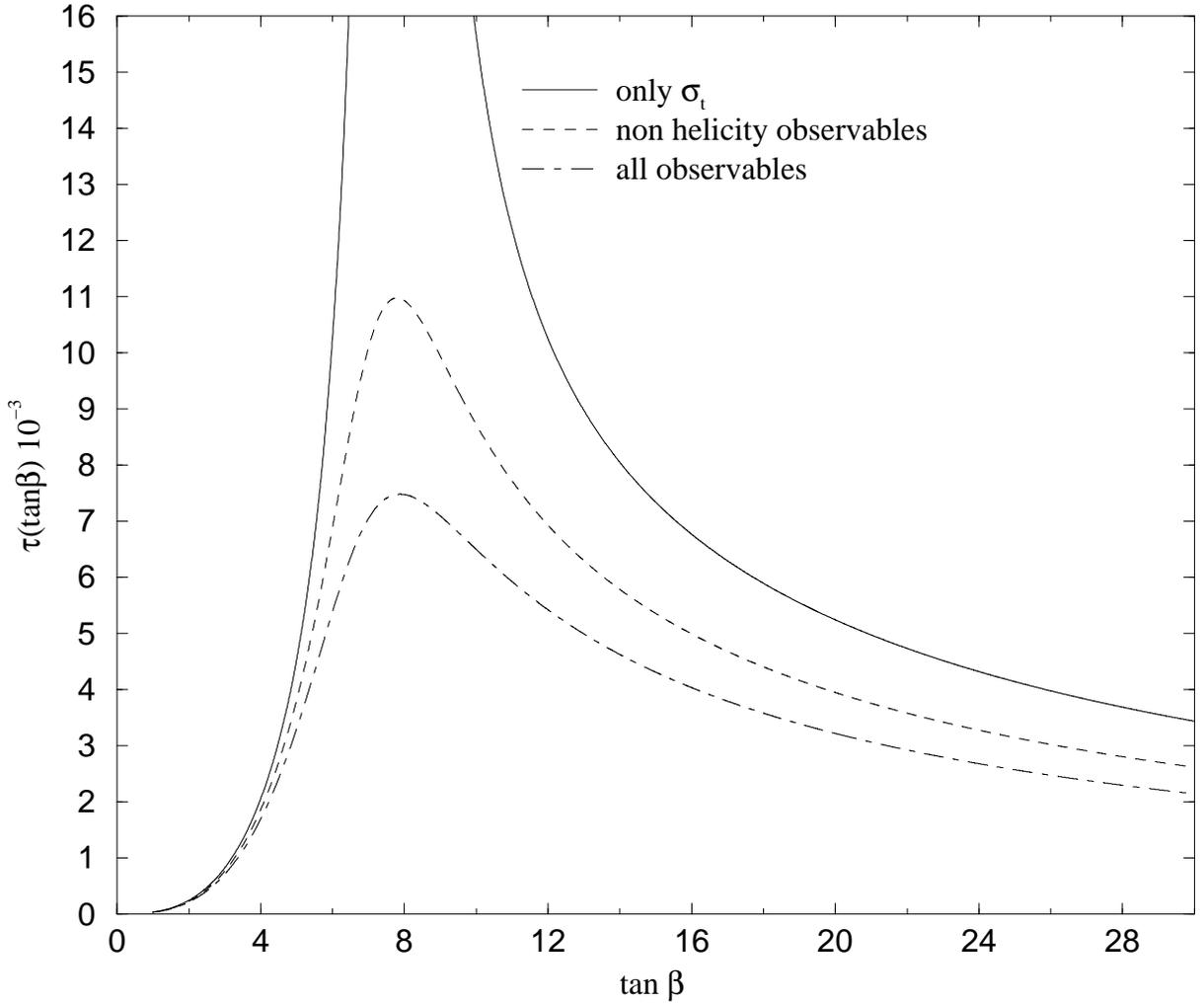,height=16cm,angle=-90}
\]
\vspace*{1cm}
\caption[1]{
Plot of the function $\tau(\tan\beta)$ when the set of observables
combined for the analysis 
is (i) $\sigma_t$ alone, (ii)  $\sigma_t$,  $A_{FB, t}$, $A_{LR, t}$ and  
$A_t$, 
(iii) the previous four and also the three helicity observables
$H_t$, $H_{FB, t}$ and $H^{LR}_t$.
}
\label{tau}
\end{figure}
\newpage

\begin{figure}[htb]
\vspace*{2cm}
\[
\epsfig{file=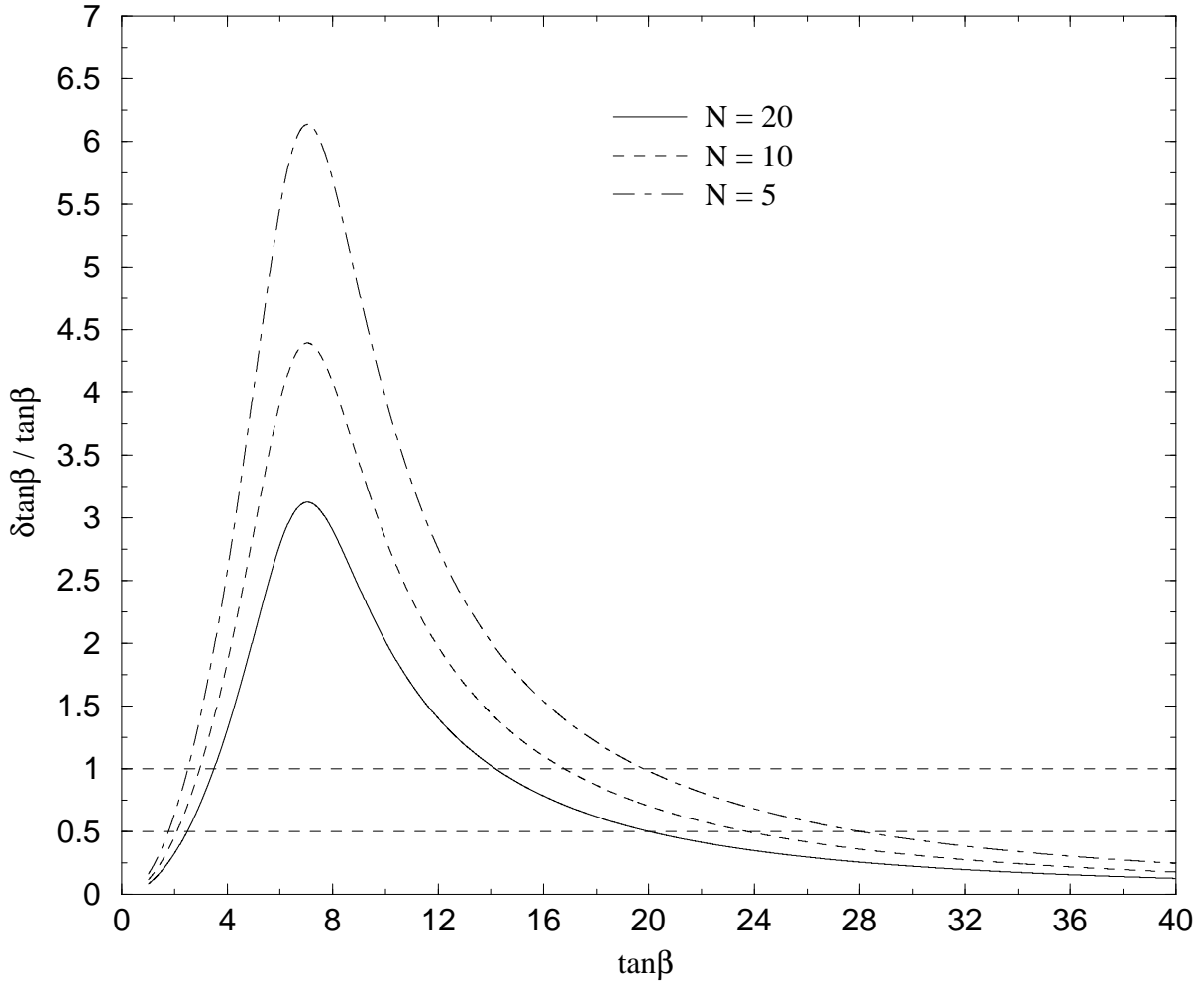,height=16cm,angle=-90}
\]
\vspace*{1cm}
\caption[1]{
Plot of the relative error on $\tan\beta$. The statistical accuracy is
$\sigma =$ 1\% for all the observables. $N$ is the number of c.m. energy values
at which independent measurements are taken. The curves actually
depend on the combination $\sigma/\sqrt{N}$.
One of the dashed lines 
corresponds to 
a 100 \% relative error. With $N=10$ measurements,  
it determines a region $3 < \tan\beta < 17$
where the determination of $\tan\beta$ is completely unsatisfactory due to the 
flatness of the coefficient of the SUSY Sudakov logarithms with 
respect to $\tan\beta$. The second line marks the $50\%$ accuracy level
and identify the region $\tan\beta < 2$ or $\tan\beta > 20$.
}
\label{error}
\end{figure}
\newpage

\end{document}